\documentclass[twocolumn,amsmath,amssymb,aps]{revtex4}

\usepackage[dvips]{color}
\usepackage{graphicx,graphics}
\usepackage{dcolumn}
\usepackage{bm}
\usepackage{amssymb}
\usepackage{slashed}
\usepackage{amsmath}
\usepackage{psfrag}
\usepackage{float}
\usepackage{pdfpages}
\graphicspath{{./Figures/}}
\def\bea{\begin{eqnarray}}
\def\eea{\end{eqnarray}}
\def\beas{\begin{eqnarray*}}
\def\eeas{\end{eqnarray*}}
\def\be{\begin{equation}}
\def\ee{\end{equation}}
\def\bes{\begin{equation*}}
\def\ees{\end{equation*}}
\def\lsim{\raise0.3ex\hbox{$\;<$\kern-0.75em\raise-1.1ex\hbox{$\sim\;$}}}
\def\gsim{\raise0.3ex\hbox{$\;>$\kern-0.75em\raise-1.1ex\hbox{$\sim\;$}}}

\def\bt{\begin{table}}
\def\et{\end{table}}

\begin{document}
\title{Light Chargino Effects onto $H \to \gamma\gamma$ in the MSSM$^\dagger$}
\author{M. Hemeda$^{1,2}$, S. Khalil$^{1,2}$, S. Moretti$^{3,4}$}
\affiliation{$^1$Center for Theoretical Physics,~Zewail City of Science and Technology,~Sheikh Zayed,~12588,~Giza,~Egypt.\\
$^2$Department of Mathematics,~Faculty of Science,~Ain Shams University,~11566,~Cairo,~Egypt.\\
$^3$School of Physics \& Astronomy, University of Southampton, Highfield, Southampton SO17 1BJ, UK.\\
$^4$Particle Physics Department, Rutherford Appleton Laboratory, Chilton, Didcot, Oxon OX11 0QX, UK.}

\begin{abstract}
We analyse the implications of light charginos on the Higgs boson signal strength via gluon-gluon fusion and di-photon
decay in the Minimal Supersymmetric Standard Model (MSSM) at the Large Hadron Collider (LHC). We show that enhancements are possible with a rate up to $25\%$. We also prove that they are possible for a high scale constrained version of the MSSM with non-universal Higgs and gaugino masses. In contrast, effects due to light charged Higgs bosons, that we also have investigated, are generically negligible in the $\gamma\gamma$ decay, though they may affect the
$b\bar b$ rate, hence the total width.
\end{abstract}
\maketitle
\hrule
\vspace*{0.25cm}
\noindent
{$^\dagger$We dedicate this work to the  lasting memory of Ahmed Elsayed.}
\vspace*{0.25cm}
\hrule
\vspace*{0.25cm}
The most recent results reported by ATLAS~\cite{ATLAS:2012au,Aad:2012tfa,ATLASNOTE2:2013,ATLAS-CONF-2013-072} and CMS~\cite{Chatrchyan:2012ufa, :2013tq, Chasco:2013pwa, CMS-PAS-HIG-13-005} confirmed a
Higgs boson discovery with a mass of order 125 GeV.  The decay channels investigated experimentally with highest
precision are
$H \to \gamma \gamma $, $H\to ZZ^{(*)} \to 4 l$ and $H \to WW^{(*)} \to l \nu l \nu$, where $l$ denotes a lepton
(electron and/or muon) and $\nu$ its associated neutrino. The data analyses in these channels are
based on an integrated luminosity of
$4.7$ fb$^{-1}$ at $\sqrt{s} =7$ TeV plus $13$ fb$^{-1}$ at $\sqrt{s}=8$ TeV (ATLAS) and $5.1$ fb$^{-1}$ at $\sqrt{s} =7$ TeV plus $19.6$ fb$^{-1}$ at $\sqrt{s}=8$ TeV (CMS). The results reported by ATLAS for the signal strengths of these  channels are given by \cite{ATLAS:2012au,Aad:2012tfa,ATLASNOTE2:2013,ATLAS-CONF-2013-072}:%
\bea%
\mu_{\gamma\gamma} &=& 1.65 \pm 0.35,\nonumber\\
\mu_{ZZ} &=& 1.7 \pm 0.5, \\
\mu_{WW} &=& 1.01 \pm 0.31.\nonumber%
\eea%
From the CMS collaboration one has instead \cite{Chatrchyan:2012ufa, :2013tq, Chasco:2013pwa, CMS-PAS-HIG-13-005}:
\bea%
\mu_{\gamma\gamma} &=& 0.78 \pm 0.28,\nonumber\\
\mu_{_{ZZ}} &=&  0.91^{+0.3}_{-0.24}, \\
\mu_{_{WW}} &=& 0.76 \pm 0.21.\nonumber%
\eea%
These results indicate suppression or enhancement in the di-photon
mode, with respect to the Standard Model (SM), with more than $2\sigma$ deviation either way, a trend which could then be
a very important signal for possible new physics Beyond the SM (BSM) \cite{Haber:1984rc,Gunion:1984yn} such as the MSSM \cite{Arbey:2012bp,Bechtle:2012jw,SchmidtHoberg:2012ip,Drees:2012fb,Arbey:2012dq,SchmidtHoberg:2012yy,Carena:2012xa,
Carena:2011aa,Hall:2011aa,Heinemeyer:2011aa,Arbey:2011ab,Draper:2011aa,Chen:2012wz,He:2011gc,Djouadi:2011aa,
Cheung:2011nv,Batell:2011pz,Christensen:2012ei}(also the constrained version \cite{Kadastik:2011aa,Baer:2011ab,
Aparicio:2012iw,Ellis:2012aa,Baer:2012uya,Cao:2011sn}), Next-to-MSSM \cite{King:2012tr,Gunion:2012gc,Belanger:2012tt,
Gunion:2012zd,Ellwanger:2012ke,Ellwanger:2011aa,Cao:2012fz,Kang:2012sy} and (B-L)SSM \cite{Elsayed:2011de,Basso:2012tr,Khalil:2012gs,
Khalil:2013in}.

In the MSSM, the Higgs sector consists of five scalar
Higgs bosons: two CP-even neutral ones, $h, H$ (with increasing mass, $m_h< m_H$), a
pseudoscalar one, $A$, and a pair of charged ones,
$H^{\pm}$. The mixing between the two CP-even neutral Higgs bosons
is defined by the mixing angle $\alpha$, which is a derived parameter. In fact, in the MSSM at the tree level, all Higgs sector observables can be defined in terms of only two input parameters, i.e., the ratio of the Vacuum Expectation Values (VEVs)
of the two Higgs doublets pertaining to this minimal realisation of Supersymmetry (SUSY), denoted by $\tan\beta$, and any of the Higgs boson masses, e.g., $m_h$.

However, in the MSSM in higher orders, genuine SUSY effects affect observables from the Higgs sector. In particular,
the mass of the lightest CP-even neutral MSSM Higgs state, $h$, typically
the SM-like Higgs, is predicted to be less than 135 GeV \cite{Djouadi:2005gj,Carena:2000yx}, owing to SUSY states entering
the one- and two-loop corrections to it. Therefore, in some sense, the
new LHC results are in favour of a low energy SUSY scenario, indeed (possibly) the MSSM.
The signal strength of the di-photon channel, $H \to \gamma\gamma$, relative to the SM
expectation, in terms of production cross section ($\sigma$) and decay Branching Ratio (BR), is defined as
\bea%
\!\!\mu_{\gamma\gamma}\!\!\!&\!=\!\!&\! \frac{\sigma(pp \to h\to \gamma\gamma)}{\sigma(pp \to h \to
\gamma\gamma)^{\rm{SM}}}\!=\!\frac{\sigma(pp \to h)}{\sigma(pp \to
h)^{\rm{SM}}}\frac{{\rm BR}(h\to \gamma\gamma)}{{\rm BR}(h \to \gamma\gamma)^{\rm{SM}}} \nonumber\\
&=& \frac{\Gamma(h \to gg)}{\Gamma(h \to gg)^{\rm{SM}}}
~\frac{\Gamma_{\rm{tot}}^{\rm{SM}}}{\Gamma_{\rm{tot}}}\frac{\Gamma(h \to \gamma
\gamma)}{\Gamma(h \to \gamma\gamma)^{\rm{SM}}} = \kappa_{gg} \kappa_{\rm{tot}}^{-1} \kappa_{\gamma\gamma}.%
\eea%

In the MSSM, the $H \to \gamma \gamma$ decay can be mediated at one loop by the $W^\pm$-boson, top quark, light squarks (in particular sbottoms and stops), light sleptons (in particular staus), charginos and charged Higgs boson. Therefore,
in such a model, the decay rate of $H\rightarrow\gamma\gamma$ is given by
\begin{widetext}
\bea
 \Gamma (h \longrightarrow \gamma\gamma) &= \frac{G_{F} \alpha^2 M_{h}^3}{128 \sqrt{2}\pi^3} \left\lvert
\sum_{\substack{f}} N_{c} Q_f^2 g_{hff} A_{1/2}^h(\tau_f) + g_{hVV} A_1^h(\tau_W) +\frac{M_W^2 \lambda_{hH^{+}H^{-}}}
{2 c_W^2 M_{H^{\pm}}^2} A_{0}^h ({\tau}_{H^{\pm}})
\right.\nonumber\\&\qquad \left. \vphantom{\sum_{\substack {f}}}+
\sum_{\substack{{\tilde\chi}_i^{\pm}}} \frac{2M_W}{m_{{\tilde\chi}_i^{\pm}}} g_{h{\tilde\chi}_i^+ {\tilde\chi}_i^-} A_{1/2}^h(\tau_{{\tilde\chi}_i^{\pm}}) +
\sum_{\substack{\tilde{f}_i}} \frac{g_{h\tilde{f}_i\tilde{f}_i}}{m_{\tilde{f}_i^2}} N_c Q_{\tilde{f}_i}^2 A_0^h(\tau_{\tilde{f}_i})
 \right\lvert^2,
\eea
\end{widetext}
where $G_F$ is the Fermi constant, $V$, $f$, and $\tilde{f}$ refer to vector, fermion and scalar particles, respectively, entering a triangle diagram.
The dimensionless parameter $\tau_i$ is defined as $\tau_i = M_h^2/4M_i^2$ with $i = f,W, H^\pm, {\tilde\chi}^\pm, \tilde{f}$ and the loop functions $A_{0,1/2,1}$ can be found in \cite{Carena:2012xa,Djouadi:2005gj}.
In Ref.  \cite{Belyaev:2013rza} a comprehensive analysis for the generic MSSM effects on the Higgs decay to di-photons, Higgs   production via gluon-gluon fusion and
total Higgs decay width was presented. In particular, it was focused on the specific effects of light stops, sbottoms
and/or staus \cite{Carena:2012gp,Giudice:2012pf,Basak:2013eba}. In this letter, we complete that analysis by revisiting chargino and charged Higgs boson effects onto,
essentially, the
$H\to \gamma\gamma$ decay rate. In fact, recall that neither $H^\pm$ nor ${\tilde\chi}^\pm_i$ ($i=1,2$) states can enter the
Higgs production mode (via $gg$-fusion) and notice that their contribution to the total Higgs boson width is subleading,
as $m_h<2M_{{\tilde\chi}^\pm_i},2m_{H^\pm}$. Therefore, in this case, one finds that $\mu_{\gamma \gamma} = \kappa_{\gamma \gamma}$.

It is worth noting that the ratio among the loop functions of a vector, fermion and scalar for masses of order ${\cal O}({100})$ GeV is about $8:1.5:0.4$. Therefore, it is clear that charged Higgs boson effects onto $\Gamma(h \to \gamma \gamma)$ are quite limited unless one can obtain a huge coupling with the SM-like Higgs via the vertex $hH^+H^-$. In contrast, chargino effects can be relevant and lead to a significant enhancement of $\Gamma(h\to \gamma\gamma)$.

\begin{figure}[t]
\begin{center}
\includegraphics[scale=0.8]{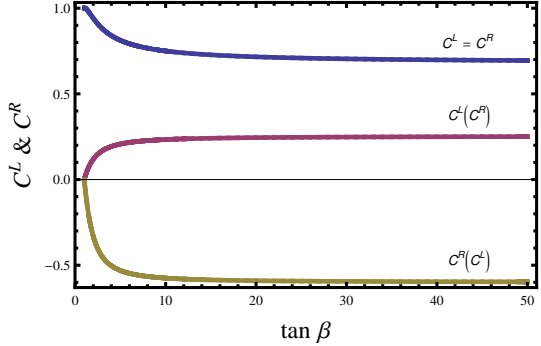}
\caption{Chargino-Higgs couplings $C_L$ and $C_R$ versus $\tan \beta$ for $\mu =M_2=200$ GeV (green line), $M_2=2\mu=300$ (red line for $C_L$ and blue line for $C_R$) and $\mu =2 M_2 = 300$ GeV (red line for $C_R$ and blue line for $C_L$).}
\label{couplings}
\end{center}
\end{figure}

In the MSSM, the chargino mass matrix is given by%
\begin{eqnarray}
{\cal M}_C = \left[ \begin{array}{cc} M_2 & \sqrt{2}M_W s_\beta
\\ \sqrt{2}M_W c_\beta & -\mu \end{array} \right],
\end{eqnarray}
where $M_2$ is the soft SUSY-breaking mass of the gaugino partner
of the $W^{\pm}$ gauge boson, the wino $\tilde{W}^{\pm}$, $\mu$ is the Higgs mixing parameter and $s_\beta \equiv \sin \beta \, , \, c_\beta \equiv \cos \beta$.
The chargino mass matrix can be diagonalised by two unitary matrices $U$ and $V$,
where $U = {\cal O}_-$ and $V = {\cal O}_+$ or $\sigma_3  {\cal O}_+$ if $\det{\cal M}_C >0$ or $\det{\cal M}_C <0$, respectively.
${\cal O}_\pm$ are rotation matrices with angles $\phi_\pm$ defined by%
\bea%
\tan2\phi_{-}&=&2\sqrt{2}M_W\frac{-\mu \sin\beta+M_2
\cos\beta}{M^2_2-\mu^2-2 M^2_W
  \cos2\beta},\\%
\tan2\phi_{+}&=&2\sqrt{2}M_W\frac{-\mu \cos\beta+M_2
\sin\beta}{M^2_2-\mu^2-2 M^2_W
\cos2\beta}.%
\eea %
The matrix ${\cal M}_C$ has two eginstates, $\tilde{\chi}_1^{\pm}$
and $\tilde{\chi}_2^{\pm}$ (the charginos), with the following mass
eigenvalues
\begin{widetext}
\be
M_{2,1}^2 =\frac{1}{2}( M_2^2 +\mu^2 +2M_W^2)\pm \sqrt{(M_2^2-\mu^2)^2 + 4
M_W^4\cos^2 2\beta + 4M_W^2 (M_2^2 +\mu^2 - 2M_2\mu \sin
2\beta)}).
\ee
\end{widetext}
The lightest chargino $\tilde{\chi}_1^{\pm}$ is often of order
$M_Z$ and has the characteristic of being the lightest charged SUSY
particle.

The interactions of the lightest neutral MSSM Higgs boson  with SM-like couplings, $h$, with the charginos are given by%
\be%
{\cal L} = g \tilde{\chi}_i^+ \left(C^L_{ij} P_L + C^R_{ij} P_R \right) \tilde{\chi}_j^+ h + h.c. %
\ee%
where the $C^{L,R}_{ij}$'s are given by%
\bea%
C^L_{ij} &=& \frac{1}{\sqrt{2}sw}\left[-\sin \alpha V_{j1} U_{i2} + \cos \alpha V_{j2} U_{i1} \right],\\
C^R_{ij} &=& \frac{1}{\sqrt{2}sw}\left[-\sin \alpha V_{i1} U_{j2} + \cos \alpha V_{i2} U_{j1} \right].%
\eea%

In Fig.\ref{couplings} we display these couplings as a function of $\tan \beta$ for different values of $\mu$ and $M_2$. As can be seen from this plot, such couplings can reach their maximum values and become of order ${\cal O}(\pm 1)$ if $\tan \beta$ is very small, close to 1, and $\mu \simeq M_2$. It is also remarkable that, if $\mu > M_2$ ($\mu < M_2$), the coupling $C_R (C_L)$ flips its sign, which leads to destructive interference between the chargino contributions. From this plot, it is clear that the Higgs coupling to charginos can be negative, hence the chargino can give a constructive interference with the $W^\pm$-boson  that may lead to a possible enhancement for $\kappa_{\gamma\gamma}$ and $\mu_{\gamma\gamma}$.

\begin{figure}[t]
\begin{center}
\includegraphics[scale=0.4]{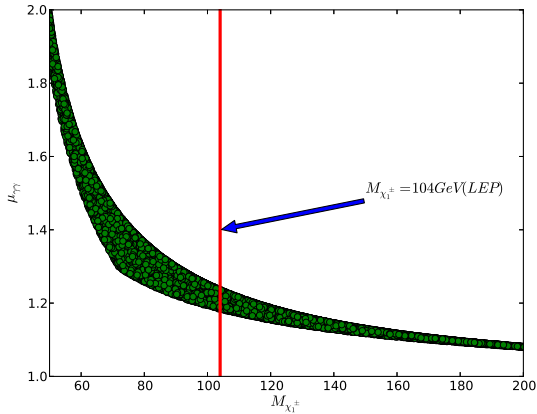}
\caption{Signal strength of the di-photon channel as a function of the lightest chargino mass for $1.1 < \tan \beta < 5$,
100 GeV $< \mu < 500$ GeV and 100 GeV $< M_2 <$ 500 GeV.}
\label{RAA}
\end{center}
\end{figure}

In Fig. \ref{RAA} we display the results for $\kappa_{\gamma\gamma}$ as a function of the lightest chargino mass, $M_{\tilde{\chi}_1^+}$, with $m_h \simeq 125$ GeV. We scan over the following expanse of parameter space (using CPsuperH (version 2.3) \cite{Lee:2003nta,Lee:2007gn}):
$1.1 < \tan \beta < 5$, $100$~ GeV $< \mu < 300$~ GeV and $100$~ GeV $< M_2 < 300$~GeV. Other dimensionful SUSY parameters
 are fixed to be of order few TeV so that all other possible SUSY effects onto $H\to \gamma \gamma$ are essentially negligible. As can be seen from this figure,  in order to have a significant chargino contribution to $\kappa_{\gamma\gamma}$, quite a light chargino mass ($M_{\tilde{\chi}_1^+} \sim 104$ GeV), around the LEP limit, is required \cite{Abbiendi:2003sc,Batell:2013bka,Haber1991,Haber:1996fp,Ellis1990}. Precisely at this limiting value, one finds that the Higgs signal strength is enhanced by about $25\%$.

As shown, the chargino mass is determined by $M_2$, $\mu$ and $\tan\beta$ and light charginos require small $M_2$ and $\mu$, which implies that $M_{1/2}$ and $m_0$ (in the case of a constrained MSSM, wherein the former/latter
represents the universal fermion/scalar mass) should be quite small. However, a Higgs mass of order 125 GeV requires quite a large stop mass $m_{\tilde t_1}$ and trilinear term $A_t$, which leads to a very large $M_{1/2}$ and $m_0$. In order to overcome this contradiction, a departure from the constrained MSSM is necessary. In particular, one has to consider non-universal gaugino masses so that $M_2$ can be much smaller than $M_3$. In addition, a non-universal Higgs mass is also crucial to guarantee small values of the $\mu$ parameter. Therefore, the following set of soft SUSY breaking terms at high scale are favoured for this analysis:
\bea
m_{H_1}^2 &=&  m_0^2(1+d_1),\\
m_{H_2}^2 &=& m_0^2(1+d_2), \\
m_{\tilde{q},\tilde{l}}^2 &=& m_0^2,
\eea%
in addition to %
\bea
M_1 &\lsim & M_2 \ll M_3, \\
A_0 & \gsim& {\cal O}(1~ {\rm TeV}).
\eea

\begin{figure}[t]
\begin{center}
\includegraphics[scale=0.8]{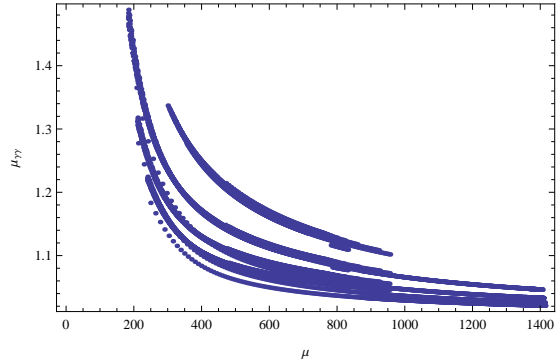}
\caption{\label{datamu} Signal strength of the di-photon channel as a function of $\mu$ for $3 \leq \tan \beta \leq 30$, $ 0 \leq d_2 \leq 5$ and 150 GeV $< M_2 <$ 250 GeV.}
\label{tanb31}
\end{center}
\end{figure}

\begin{figure}[t]
\begin{center}
\includegraphics[scale=0.8]{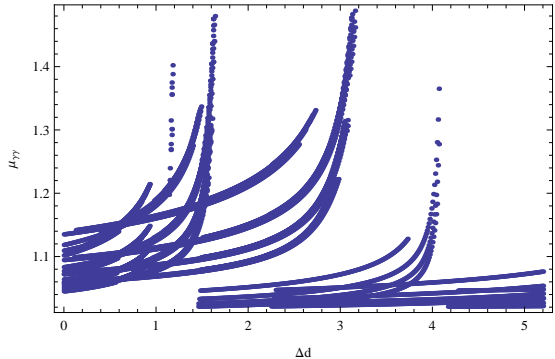}
\caption{\label{data1} Signal strength of the di-photon channel as a function of the difference $\Delta d = d_2 - d_1$
for $3 \leq \tan \beta \leq 30$, $d_1 = -1$, $ 0 \leq d_2 \leq 5$ and 150 GeV $< M_2 <$ 250 GeV.}
\label{tanb5}
\end{center}
\end{figure}

Running these soft terms from the Grand Unification Theory (GUT) scale down to the SUSY scale $\sim \sqrt{m_{\tilde{t}_1} m_{\tilde{t}_2}}$ and
imposing the electroweak breaking conditions, one finds that the $\mu$-parameter is given by%
\begin{eqnarray}
\mu^2=\frac{m_{H_1}^2 - m_{H_2}^2 \tan^2\beta}{1-\tan^2\beta}-\frac{M_Z^2}{2}.
\end{eqnarray}
One can easily show that $\mu$ is strongly dependent upon the value of $d_1$ and $d_2$ and that for $d_1 \ll d_2$ a light $\mu \sim {\cal O}(100$ GeV) is achieved, so that
we obtain different values for $m_{H_1}$ and $m_{H_2}$ in correspondence to a small $\mu$. In Fig. \ref{datamu} we display the results for $\mu_{\gamma\gamma}$ versus $\mu$ for $m_0 \sim A_0 \sim 1$ TeV and $ 500 \leq M_3 \leq 1100$ (so that $m_h \approx 125$ GeV), $3 \leq \tan \beta \leq  30$, $ 0 \leq d_2 \leq 5$ and 150 GeV $ < M_2 <$ 250 GeV.
This figure confirms that a quite small $\mu$ is obtainable in this class of SUSY models and the signal strength in $ \gamma \gamma$ is significantly enhanced for these values.  In Fig. \ref{data1} we display the results for $\mu_{\gamma\gamma}$ as a function of the difference $\Delta d = d_2 - d_1$. Here, we vary the other parameters in the aforementioned regions. As can be seen from this plot, for $\Delta d > 1$ the signal strength can be enhanced and become larger than one. Also, it may have a resonant behaviour in the regions where $\mu \sim M_2$.

\begin{figure}[!t]
\begin{center}
\includegraphics[scale=0.4]{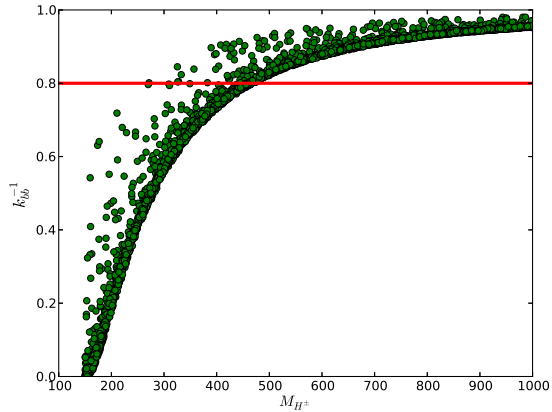}
\caption{$\kappa_{bb}^{-1}\simeq \kappa_{\rm tot}^{-1}$ versus the charged Higgs boson mass. Here,
$1.1 < \tan \beta < 50$,
$\mu \sim M_2 \simeq 2000$ GeV and 100 GeV$<~m_{H^\pm} <1000$ GeV.
Further, the red line represents the lower limit on $\kappa_{bb}^{-1}$ roughly compatible with current LHC data.}
\label{kbb}
\end{center}
\end{figure}

Finally, as anticipated, a numerical analysis confirmed that the charged Higgs boson contribution is generally negligible
in the $h\to\gamma\gamma$ decay, so we do not produce the corresponding formulae.
However, we do present here an interesting plot, highlighting the key role of the $H^\pm$ mass entering the $hb\bar b$ coupling squared (but not the effective $h\gamma\gamma$ one), see Fig. \ref{kbb}, wherein $\kappa_{bb}\equiv\Gamma (h\to b\bar b)/\Gamma(h\to b\bar b)^{\rm SM}$  (also recall that
$\Gamma_{\rm tot}\approx\Gamma (h\to b\bar b)$).
In this figure, using again CPsuperH (version 2.3) \cite{Lee:2003nta,Lee:2007gn}, we assume that $100$ GeV $<~m_{H^\pm}<1000$ GeV,
$\mu \sim M_2 \simeq 2000$ GeV (so that chargino effects are completely decoupled). In fact, owing to the mass relation
between the charged and pseudoscalar Higgs bosons, i.e., $m_{H^\pm}^2 = m_A^2 +m_{W^\pm}^2$ (at tree level), this argument can be recast in terms of $m_A$.
The point is that the MSSM rescaling factor of the $hb\bar b$ coupling (at tree level) is $\sin\alpha/\cos\beta$ and that the $\alpha$ and $\beta$ angles are related via
the well-known (tree level) formula
\begin{equation}
\tan 2 \alpha =\tan 2 \beta ~\frac{m_{H^\pm}^2-m_{W^\pm}^2 + m_Z^2}{ m_{H^\pm}^2-m_{W^\pm}^2 - m_Z^2},
\end{equation}
so that there exists a strong correlation between  $\kappa_{bb}^{-1}$ and $m_{H^\pm}$: in particular,
the smaller $m_{H^\pm}$  (or $m_A$) the smaller $\kappa_{bb}^{-1}$.  This is well exemplified by noting that the  edge of the distribution of green points in
Fig. \ref{kbb} is nothing but $(\sin\alpha/\cos\beta)^{-2}$, with the spread determined by the actual value of $\tan\beta$
(and subleading loop effects). It is therefore clear the potential that a measurement of
$\kappa_{bb}$ can have in (indirectly) constraining $m_{H^\pm}$  (or $m_A$),
even in the region presently compatible with LHC data (above the red line). We find such effects to be generally realised also
in the constrained version of the MSSM.

In conclusion, we have proven that both chargino and charged Higgs effects induced by the MSSM can affect the LHC data
used in the Higgs search
over significant regions of the paremeter space of such a minimal SUSY realisation, including in its constrained version,
so long that non-universal gaugino and Higgs masses are allowed. Light charginos can increase significantly the
$h\gamma\gamma$ (effective) coupling whereas light charged Higgs bosons can sizeably increase the $hb\bar b$ one.
Whereas the former effect could easily be confirmed or disproved by upcoming LHC data (at higher energy and luminosity) by
measuring the $\gamma\gamma$ signal strength, the latter phenomenon may be more difficult to extract via the
$b\bar b$ signal strength, as the $h\to b\bar b$ partial decay width is
very close to the total one. Finally notice that such ${\tilde\chi}^\pm_i$ and $H^\pm$ effects are normally realised on non-overlapping regions of parameter space, so that they would not appear simultaneously.

\section*{Acknowledgements} The work of SM is supported in part through the NExT Institute. The work of MH and SK  is partially supported by the ICTP grant AC-80. We are grateful to Sasha Belyaev and Marc Thomas for insightful discussions
and for numerical tests.

\bibliographystyle{h-elsevier}\bibliography{draft}
\end{document}